*Long Article*

# A Novel Approach for Network Attack Classification Based on Sequential Questions

**Md Mehedi Hassan Onik[1,*], Nasr Al-Zaben[1], Hung Phan Hoo[2], Chul-Soo Kim[1]**

[1]Department of Computer Engineering, Inje University, Gimhae-50834, Republic of Korea
hassan@oasis.inje.ac.kr, Nasr.zaben@hotmail.com, charles@inje.ac.kr
[2]Team lead & Android Developer, True Corporation, Bangkok-10310, Thailand
hungphan88@gmail.com
*Correspondence: hassan@oasis.inje.ac.kr

**Abstract:** With the development of incipient technologies, user devices becoming more exposed and ill-used by foes. In upcoming decades, traditional security measures will not be sufficient enough to handle this huge threat towards distributed hardware and software. Lack of standard network attack taxonomy has become an indispensable dispute on developing a clear understanding about the attacks in order to have an operative protection mechanism. Present attack categorization techniques protect a specific group of threat which has either messed the entire taxonomy structure or ambiguous when one network attacks get blended with few others attacks. Hence, this raises concerns about developing a common and general purpose taxonomy. In this study, a sequential question-answer based model of categorization is proposed. In this article, an intrusion detection framework and threat grouping schema are proposed on the basis of four sequential questions ("Who", "Where", "How" and "What"). We have used our method for classifying traditional network attacks in order to identify initiator, source, attack style and seriousness of an attack. Another focus of the paper is to provide a preventive list of actions for network administrator as a guideline to reduce overall attack consequence. Recommended taxonomy is designed to detect common attacks rather than any particular type of attack which can have a practical effect in real life attack classification. From the analysis of the classifications obtained from few infamous attacks, it is obvious that the proposed system holds certain benefits related to the prevailing taxonomies. Future research directions have also been well acknowledged.

## 1. Introduction

Network attack classification is the process of grouping network attacks to specific subgroups in order to determine similar types of attack in future. The purpose of this classification is that it can help us to know more detail about the network attack characteristics like origins, scopes, initiator and seriousness of an attack. We can also plan effective defences and preventive measures as well to reduce attacks' consequences for global networks. Network attack classification is the first step to have a clear idea about attacking style and subsequent system protection.

The fast increasing of network attacks in both scales and severities encourage us to classify and investigate in detail about the network attacks. There are many research on base of network attack classification. Vulnerabilities [1-2], lists of term taxonomy [3], application of taxonomy [4-7] and multiple dimensional taxonomies [8-9] etc. are important. Before defining a classification for network attacks, it is important to define the requirements which must be compiled with the new classification [10]. Bailey with Bishop in a study, outline a classification which lets exclusive identification of objects [3]. Categorizations of attack served as a helpful tool in modelling security guidelines for a defence mechanism. Here, we selected some requirements that relevant to the proposed classification by studies [10-11]: **Accepted** [12]: The taxonomy can be generally approved. The taxonomy must be designed so that it becomes commonly accepted one. **Understandable** [12]: Classification should be easy to understand by those who are in network, security or related field. **Completeness** [13]: In order for a classification to be complete, all network attacks must be included in this classification and have a specific category. It is difficult to prove a classification has accomplished, but it could be accepted based on the successful categorization of the actual attacks. Above two reflect that a taxonomy should be accountable for all threat and capable of categorizing them. The classification should be acceptable through successful categorization of the threats. **Mutually exclusive** [13]: This requirement categorizes each assault into one class. **Repeatable** [12-13]: Classification needed to be repeatable. **Unambiguous** [12-13]: Grouping must be defined clearly in such a way that there is no doubt as to what category the network attack should be in. **Useful** [12-13]: A useful classification could be used in the network field, or security field, or other related fields.

Other early taxonomies were Protection Analysis (PA) plus Research in Secured Operating System (RIOS) [14-15]. They also focus on vulnerabilities rather than attack, but they provided the categories on security defects and lead to related grouping arrangements. Direct use of syntax and semantic relations between attacks by ontologies were discussed in the study [15]. Field-specific taxonomies are there like for computer worms [16] and standardized attack [15]. Several of these taxonomies will be introduced in the next part of this study. Hidden Markov model (HMM) with Markov model is currently being used for attack classification. HTTP payload was analysed in work like this with HMM [17]. Pattern identification is the main idea of the attack categorization. HMM can easily define unknown parameters, through the observation and feature considerations. Healthcare and financial anomaly detection were mentioned by a study like this [18]. Where false data injection attacks [FDIA] with their full scope on smart grid and healthcare technology were discussed. Network equipment as switches, routers are also at high risk of attack. This type of attack costs money and energy losses to recover from that attack with devices was discussed by Onik [19].

This study proposes, a new approach that is constructed on the sequential questions: 'Who', 'Where', 'How' and 'What'. The attacks, which have the same type of attackers (Who), same locations where attacks were begun (Where), using some similar tools to attack (How) and degree and type of attack range (What), could be considered in the same type of attacks. Our proposed classification takes a different approach than the above classifications but also uses them as a part of our taxonomy. We create this classification based on the normal sequences when all network attacks occurred. First, all network attacks must be controlled by a person or a group of people, organizations, as well as governments. That why "Who" is the first question in our model. Second, all attacks must have a starting point from some places or locations and also have destinations to destroy or destruct. This is our "Where" question. Next "How" question covers the tools, the ways or the vulnerabilities that the network attacks could exploit to perform their actions. This is the most complicated question in our taxonomies. The last question "What" describes the intensity of the network attacks. This question could help us to determine the scale and influent scope of network attacks from its consequence.

Rest is presented as. Section 2, presents requirements for attack classification. Section 3, shows related works with advantages and disadvantages. Our proposed classification is discussed in Section 4. Section 5 compares and discusses ours' with well-known network attacks. Section 6 draws conclusions, summarizes the proposed model and feature works followed by necessary references.

## 2. The Necessity of Network Attack Classification:

However, the main reason for a new taxonomy is lack of standard and globally accepted classification. The first problem is, most of the taxonomies are related only to a specific field of interest. The second problem is, how existing study has classified the blended attacks. The attacks that contain other attacks cause a messy structure during classifications. List of those classifications would become almost infinite and there are few instances within each category or multi-dimension taxonomies where each leaf node could point to other leaf nodes and makes them difficult to be used for classification. The last problem is, existing taxonomies have to face with the unlimited sub-branches in their classifications when network attacks don't have many common traits. Therefore, the simplicity of the list classifications or the heritance of multi-dimensional ones lost. Although collective anomaly detection and their techniques for network traffic attack were analysed and discussed by few studies, still for further generalization, we proposed this sequential question-based attack classification [20]. Another classification is proposed by MIT Lincoln Laboratory where multi-dimensional assault grouping was done on the basis of the level of privileges [21]. In that classification, attacks were divided as public vs local, user vs root, investigation and denial of services (DoS) [21]. With these requirements, in the next section, we discuss some previous classifications. We outlined three reasons for a new classification by evaluating these studies [22-24].

1.  Very often, administrators find difficulty to detect the exact attack sub-group due to complex taxonomy. This caused delay and that makes the situation worst.
2.  Organizations are collecting attack information differently with their own way of classification. However, in future, those data cannot help other classification. Since our proposed classification is applicable for every kind of attacks, this taxonomy can easily use collected information in the future for a similar case.
3.  There is no fixed or standardized taxonomy, lots of taxonomies are being created with different viewpoints.

## 3. Literature Review:

***Based on vulnerability classification:*** Based on genesis, instruction time and location, Landwehr presented one of the earliest attack taxonomy that is shown in Figure. 1 [2].

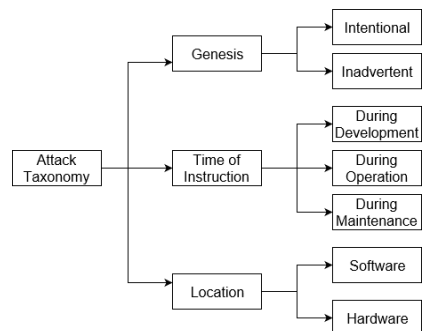

Figure 1. Attack classification by security flaw [2].

In another way, Matt Bishop [3] introduced a categorization of UNIX weaknesses in which the core faults of weaknesses were being used to make attack taxonomy. He introduced six "axes" to arrange vulnerabilities: time of initiation, nature, exploitation sector, minimum number, attacked domain and attack causes. Bishop suggested that one of the key advantages of a classification is that it ought to help working out where to invest resources to prevent an attack.

***Based on type of vulnerability classification:*** List of terms were the simple and popular taxonomy but that couldn't help much. They included a necessary longer list of attacks terms without classifying. Cohen presented terms for threat grouping: harassment, denial of services, hiding, illegal information duplication, software piracy, reduction of services quality, worms and malware etc. [1].

| | | Access required | | | |
|---|---|---|---|---|---|
| | Attacker | Possesses the Device | Handles the device | Approaches the device | Interface with the device |
| Action | Recover a Key | | | | |
| | Defeat Authentication | | | | |
| | Avoid Authentication | | | | |
| | Deny Service | | | | |

Figure 2. Attack classification matrix [4].

Secondly, Alvarez [5] proposed a web threat grouping, in which he included around ten classes and sub-classes. This research focused on the principles of the attack progression that assisted to understand the characteristics and way of attack.

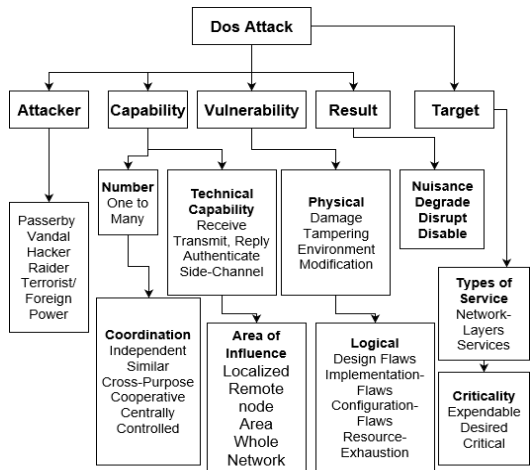

Figure 3. Classification designed for Denial-Of-Service attacks [6].

Another taxonomy related to denial-of-service (DoS) attack classifications is proposed by Anthony and Mirkovic [6-7]. These classifications are only specular for DoS attacks but it can help us to identify the attackers, his capabilities, targets, vulnerabilities and his end results. Another field, they can help us to exploit the weakness, a communication mechanism, automation degree, the impact on victims etc. Figure 3 and figure. 4 show the summary of these classifications.

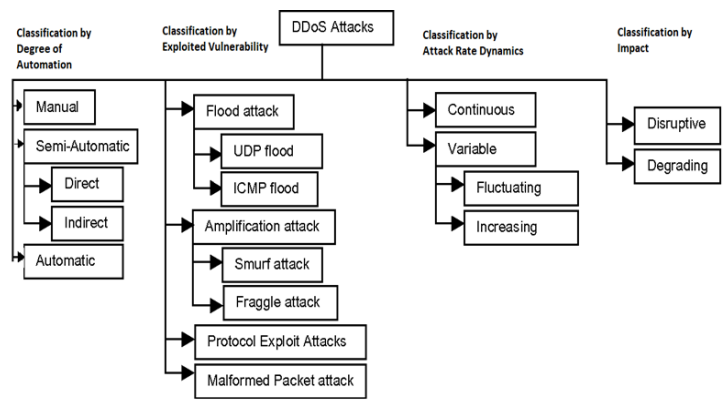

Figure 4. Classification designed for DDoS attack mechanism [7].

***Based on multiple dimension classification:*** Nowadays, describing the network attacks with the single attribute cannot cover all the processes of attack characteristics. So, there are several approaches which are based on multi-dimension classification.

In [9], Howard proposed a classification for network and computer attacks which got five stages: access, tools, objectives, attackers and results. The attackers are the types of people who launched an attack. Tools are considered as the way that attackers used for performing their actions. Access is completed by implementation, formation or design weaknesses. After the access is reached, the outcome could be theft of service or information corruption. This classification focused on attack, not on its process. Figure. 5 shows Howard's attack taxonomy.

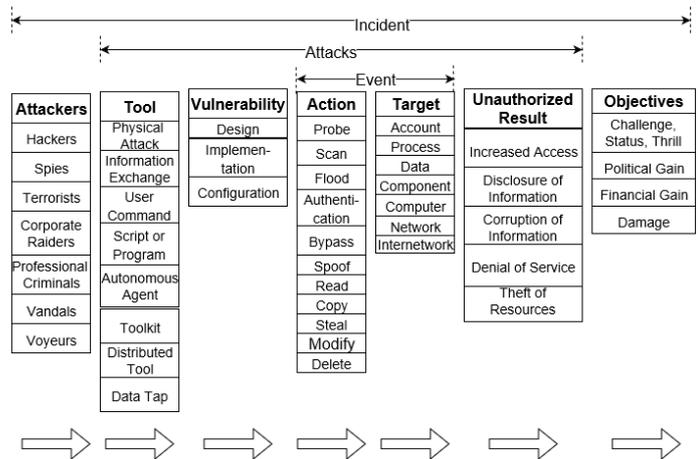

Figure 5. Howard's attack taxonomy [9].

Daniel Lough presented a classification named validation (V) exposure (E) randomness (R) deallocation (D) improper (I) conditions (C) taxonomy (T) shortly as VERDICT in 2001 on the basis of the attack features [8]. Daniel Lough used four features to describe his VERDICT. Firstly, inappropriate validation. Secondly, improper exposure. Thirdly, unsuitable randomness. Finally, inappropriate deallocation.

Hansman [10] used the concepts of dimensions to introduce his computer and network attack classification. There are four dimensions in Hansman's taxonomy. The first aspect is being used to categorize threat into a group which is created on the attack vector. The second aspect covered the attack target. The vulnerabilities and exploits are covered in the third aspect. The final aspect considers the possibility of a threat to have a payload or outcome which does not belong to itself.

## 4. Proposed Classification

***The motivation of our proposed classification:*** Our classification focuses on four sequential questions network attack processes with are: Who, Where, How and What. The approach is based on an idea that all similar network attack have a similar way to attack and the classification is built with those four questions. By following the network attack process from launching to ending, this approach can provide a better approach which adapts all requirements of a network attack classification as well as covers all current network attacks in a simple way which is very helpful for future. Four questions link together as shown in Figure. 6. Next part illustrates the detail of each question and way in which they are being used to provide a complete classification.

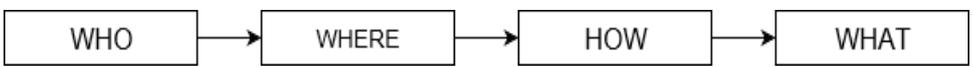

Figure 6. Four sequential questions in the proposed classification.

### 1.   *Who launched a network attack?*

Our classification focuses on four sequential questions network attack processes with are: Who, Where, How and What. The approach is based on an idea that all similar network attack have a similar way to attack and the classification is built with those four questions. By following the network attack process from launching to ending, this approach can provide a better approach which adapts all requirements of a network attack classification as well as covers all current network attacks in a simple way which is very helpful for future. Four questions link together as shown in Figure. 6. Next part illustrates the detail of each question and way in which they are being used to provide a complete classification. These five categories and their related objectives are shown in Figure. 7.

- **Joker** – perform a network attack primarily on the learning and challenges. An example can be Jonathon James was a US student hacked US department of defence and NASA. Similar cases were mentioned in this blog [25].

- **White-hat hackers** – perform a network attack to find out the vulnerabilities of the attacked network and report to the network administrator. This type of hackers just finds the backdoor for helping administrator to stop future attacks like this [26].
- **Black-hat hackers** – perform a network attack by exploiting some vulnerabilities of the network and damage or stole the information from the attacked network.
- **Little sisters** – the organizations or companies who launch attacks on competitor's network for financial gain.
- **Big brothers** – the governments or the government-related organizations launch attacks primarily in order to achieve political gain. For example, some hacking group were active after Donald trump won US elections was mentioned by this study [27].

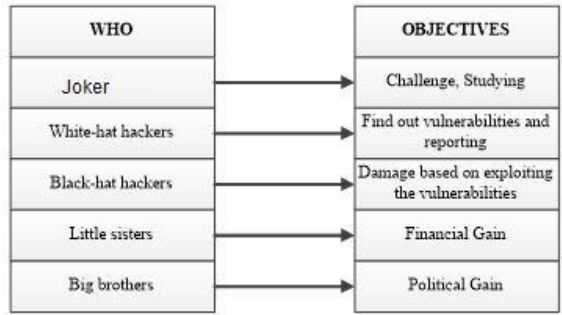

Figure 7. WHO question.

## 2. *WHERE is network attack from and WHERE is it to?*

In WHERE question, all network attacks always have the initiated points to be launched and their attack scopes are depended on the objects and the WHO in the previous question. Therefore, we divided WHO question into two sequences: (i) Initiated locations and (ii) Attack scope. The relationship between initiated location and the attack scope of WHERE is shown in Figure. 8.

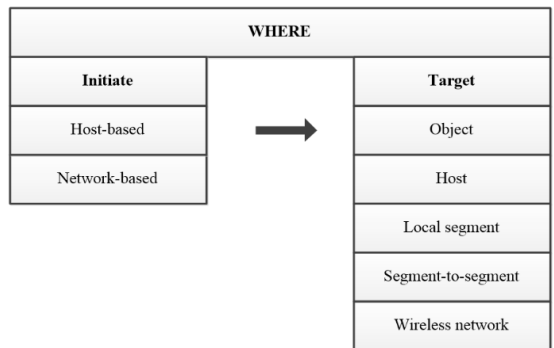

Figure 8. WHERE question.

i.  **Initiated location**: For initiated location, there are two types of address. One is host-based initiation that an attack is launched from a computer or any device that has a network connection and one is network-based initiation that an attack could be launched by multiple devices connected together.

ii. **Attack Scope**: With attack scope, we separate every network attack by five categories as:

- **Object-based** – the target of the attack is a single object in real life which has a network connection, such as a car, a mobile-phone, a smart-watch and so on. In here, we can have some groups of object: computer, mobility device, embedded device and network equipment.
- **Host-based** – the target of the attack is on a computer terminal like a personal computer, a server and after gained access on this host, the attack can be easy to expand to other hosts in the same network with the victim host.
- **Local segment-based** – the target of the attack is on a segment of the network that has many hosts connected with each other. For example. Metropolitan Area Network is one example. Other similar examples are Local Area Network as well as Wide Area Network.

- **Segment-to-segment-based** – This type of target tries to attack in the core of the global network (User-to-Network Interface, Network-to-Network Interface), for example in Border Gateway Protocol.
- **Wireless network-based** – the target of the attack is on the mobile network. Such as Bluetooth and WiFi hotspot.

3. *How does the attack succeed?*

The HOW question can be said in another way that how a network attack can perform their actions and gain the accesses from the attacked system. To answer this question, we proposed three sub-processes: Vulnerabilities, hacking tool platform and attack channel. There are already many taxonomies for the vulnerabilities. However, classifying the vulnerabilities is out of scope in this paper. In here, we focus on the way to exploit the vulnerabilities by using some hacking tool platforms and some attack channels. To perform the hacking actions quickly, the WHO should use some hacking tool platforms. In this paper, we propose four types of platform Figure. 9.

- **Software** – hacking platform that based on Operating system (OS) of devices or applications installed on devices.
- **Hardware** – hacking platform that based on devices' physical accessing to change their normal functions
- **Embedded hardware** – a hacking platform that used the firmware of devices to perform the hacking actions, as well as to change the features of firmware for attacker's purposes.
- **Mobile** – new rising hacking platform that got unauthorized permissions from applications installed on mobile devices, or from SMS/MMS services.

Using hacking tool platforms, an attacker from WHO must rely on some channels to access and to steal information. In here, five types of the channel are proposed as follows Figure. 9.

- **Legacy network equipment ports** – the type of channel that followed by standardized network protocols
- **Undefined network equipment ports** – the type of channel that followed by some special network protocols, that are produced by manufacturers.
- **Virtualization channel** – the type of channel is based on cloud computing or virtualization technique.
- **User-to-network channel** – the normal channel, which is used in daily network activities, is exploited by an attacker. Like as (MITM) Man-in-the-middle or DDoS botnet.

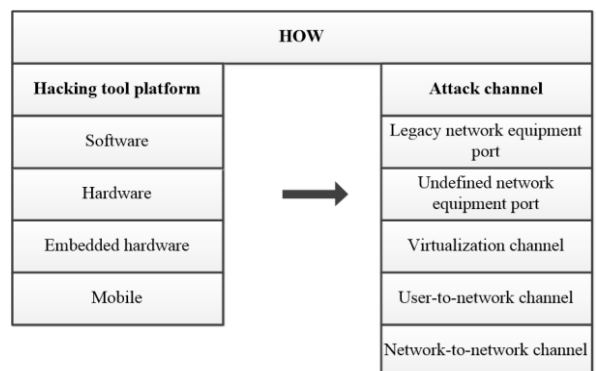

Figure 9. HOW question.

- **Network-to-network channel** – the channel is relied on in the core of the network, is exploited by using some segment-by-segment protocols.

4. *What is the type of the attack?*

The last question is about the intensity of the network attacks into the specific networks. This question belongs at the end of attack process when attackers from WHO already gained systems and can control attacked networks by themselves through WHERE and HOW. On the other hand, the intensity and type of an attack depend on the objectives of the attacker and it can be divided by three type according to WHAT question. Follows are the three situation which helps us to define the type of attack after the virus has already infected the system. This tells us to what extent we should defend.

With this WHAT question, we can detect the strength and type of the attack. Our intention is to know the class and effect of the attack with "What" question to an attack. When any one of three happen we detect that type of attack it is. Figure. 10 depicts these types of WHAT question.

- **Abnormal system activities** – when the network has some abnormal activities from it resources such as CPU utilization, disk utilization, or network utilization.
- **Traffic volume** – when the network has to face to response a number of requests to steal their information. Only restriction and limitations over data are there with this subcategory.
- **Controllable requests** – when the network is detected that occurred some abnormal requests from host-based or network-based.

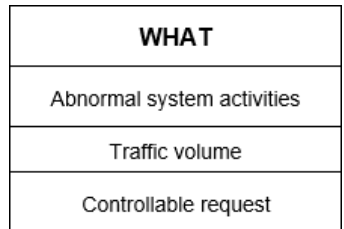

Figure 10. WHAT question.

### 5. *Overall Taxonomy in a nutshell:*

The overall taxonomy of all its subclasses is shown in below Figure. 11. This taxonomy does not focus any special sector of attack rather, it can classify every kind of attacks. An attack can easily defence if an administrator can know about the attacker, how it attacks and how much trouble a particular attack can cause with our proposed taxonomy stated below.

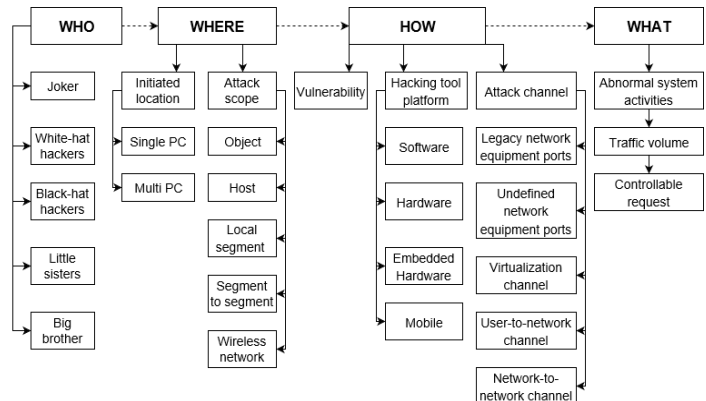

Figure 11. Sequential Question-based Network attack taxonomy.

## 5. Evaluation and Discussion

In last two section, we discussed other classification methods and proposed out classification. In this section, we will now evaluate our classification by classifying few network attacks and worms with our proposed system. Firstly, we will compare the presence of key characteristics according to a study [10] with two other studies [28-29] who did similar kind of classification shown in Table. 1.

TABLE 1
Comparison of attack taxonomy characteristics.

| Requirements | Van Heerden et al [28]. | Simmons et al [29]. | Proposed Approach - Sequential Question |
|---|---|---|---|
| Accepted | Yes | Yes | Yes |
| Comprehensible | Yes | Yes | Yes |
| Conforming | Yes | Yes | Yes |
| Determined | Yes | Yes | Yes |
| Exhaustive | No | Yes | Yes |
| Mutual Exclusion | Yes | No | Yes |
| Repeatable | Yes | Yes | Yes |
| Well Defined | No | Yes | Yes |
| Unambiguous | Yes | Yes | Yes |
| Useful | Yes | Yes | Yes |

Inside this part, the proposed taxonomy 'Sequential Attack' is being matched with the past classifications described by previous sections. Evaluation outcome shows how effectively 'Sequential Attack' catches the attacks and threats info. Also delivers countermeasures for stopping attacks.

1.  ***Blaster Attack classification:*** Windows XP along with Windows 2000 of Microsoft system were the main victims of the Blaster worm. This computer worm attacked many computer systems during August 2003 [30]. Type: Internet Worm. Creator: Unknown but people also says Jeffrey Parson. Platform and file category: Microsoft Windows and .exe. Reported cost around $320 m and source language 'C'. Classification of Blaster attack shown in Table.2.

TABLE 2
Blaster attack taxonomy.

| Classification by Lough: VERDICT [8] | | | | |
|---|---|---|---|---|
| Attack Name | Inappropriate Validation | Inappropriate Exposure | Inappropriate Randomness | Inappropriate Reallocation |
| Blaster | None (X) | None (X) | | |
| Taxonomy by Howard [9] | | | | |
| Threat Name | Attack utensils, tools | Threat Weakness | Action | Attack Goal | Unlawful outcome |
| Blaster | Computer Program | The overflow of the Buffer | Change | Computer Network | Data tampering |
| Classification by Hansman with Hunt [10] | | | | |
| Threat Name | First Dimension | Second Dimension | Third Dimension | Fourth Dimension |
| Blaster | System Network-based Worm | Network | CAN-2003-0352 | TCP and UDP overflow, DoS |
| Joshi's ADMIT Classification method [31] | | | | |
| Threat Name | Attack Vector (A) | Defence (D) | Method (M) | Impact (I) | Target (T) |
| Blaster | The overflow of the buffer | While listing patch method | System virus | Distort | MS XP and MS 2000 |
| Proposed Taxonomy: Sequential Question | | | | |
| Attack Name | Who | Where | How | What |
| Blaster | Black-hat hackers (Jeffrey Parson) | Initiated by the host a Single PC attack (already attacked PC) | Embedded legacy network equipment port (TCP port 135) | Controllable request (Can control TCP port 4444 and UDP port 69) |

2.  ***Melissa Attack classification:*** It was found on 26th of March, 1999. This affected the internet and led to shutting it down those email services that blocked in with the attacked e-mails spreading from the Melissa virus. Though the main intention of the virus was not bad, somehow it filled servers to make the situation worst [32]. The basic language of this virus was visual basic of Microsoft. Type: Word macro virus. Creator: "Kwyjibo". Platform: Microsoft Windows and Word. File Category: Docs and Doc. Reported cost: $1.1 Billion. Place of Origin: Aberdeen, New Jersey, USA. Detail classification of Melissa attack with proposed and other attack classification are shown in Table.3.

TABLE 3
Melissa attack taxonomy.

| Classification by Lough: VERDICT [8] | | | | |
|---|---|---|---|---|
| Attack Name | Inappropriate Validation | Inappropriate Exposure | Inappropriate Randomness | Inappropriate Reallocation |
| Melissa | | None (X) | | None (X) |
| Taxonomy by Howard [9] | | | | |
| Threat Name | Attack utensils, tools | Threat Weakness | Action | Attack Goal | Unlawful outcome |
| Melissa | Script | Setup | Verification | Information | Data tampering |
| Classification by Hansman with Hunt [10] | | | | |
| Threat Name | First Dimension | Second Dimension | Third Dimension | Fourth Dimension |

| Melissa | Bulk-emailing worm | MS word 97 and MS 2000 | Setup | Macro worm & TCP data packet overflow | |
|---------|--------------------|-----------------------|-------|--------------------------------------|--|
| Joshi's ADMIT Classification method [31] | | | | | |
| Threat Name | Attack Vector (A) | Defence (D) | Method (process) (M) | Impact (Effect) (I) | Target (goal) (T) |
| Melissa | Setup in a wrong way | Path system | Virus: Bulk emailing | Disrupt | App: MSW 97, 2000 |
| AVOIDIT Taxonomy [29] | | | | | |
| Attack Name | Attack Vector (AV) | Operational Impact of attack(OI) | Defence (D) | Impact of the attack (I) | Target or goal of the attack(T) |
| Melissa | Misconfiguration | Attack with email | List email addresses | Identify other ways | Microsoft products |
| Proposed Taxonomy: Sequential Question | | | | | |
| Attack Name | Who | Where | How | What | |
| Melissa | Joker (Kwyjibo) | Initiated by multiple PC of wireless media with software level hacking tool | User to network channel use which brings | Abnormal system activity | |

3. *Slammer Attack classification:* This virus also called SQLExp, Sapphire or Helkern. It was first noticed on January 25th 2003 and later become famous by becoming the fastest growing virus in all over the world during that period [33]. Type: Internet Worm. Creator: "Unknown". Basis language: Assembly. Platform: Microsoft Windows. File category: UDP packet. Reported cost: $1.2 Billion. Detail classification of Slammer attack with proposed and other attack classification are shown in Table.4.

TABLE 4
Slammer attack taxonomy.

| Classification by Lough: VERDICT [8] | | | | |
|--------------------------------------|--|--|--|--|
| Threat Name | Inappropriate Validation | Inappropriate Exposure | Inappropriate Randomness | Improper Reallocation |
| Slammer | None (X) | None (X) | | |
| Taxonomy by Howard [9] | | | | |
| Threat Name | Attack utensils, tool | Threat Weakness | Action (way of threat) | Attack Goal | Unlawful outcome |
| Slammer | Computer Script | Setup and design | Problem, change | System Network | Data corrupt |
| Taxonomy by Hansman with Hunt [10] | | | | |
| Attack Name | First Dimension | Second Dimension | Third Dimension | Fourth Dimension (degree) |
| Slammer | Computer network-Aware worm | Microsoft SQL Server 2000 | CAN-2002-0649 | Buffer run-off and UDP data flood and DoS |
| AVOIDIT Classification [29] | | | | |
| Threat Name | Attack Vector (AV) | Operational Impact (OI) | Defence (D) | Impact (I) | Target (T) |
| Slammer | Misconfiguration | Setup virus and malware: Network-based | Moderation style: Whitelist CVE- 0649 | Discover | Network |
| Joshi's ADMIT Classification [31] | | | | |
| Threat Name | Attack Vector (A) | Defence (D) | Method (Procedure) (M) | Impact (Effect) (I) | Target (Goal) (T) |
| Slammer | Wrong setup | A patch of the system | Virus: setup worm | Identification | Network |
| Proposed Taxonomy: Sequential Question | | | | |
| Attack Name | Who | Where | How | What |
| Slammer | White-hat hackers (Benny, 29A) | Single PC, Host base attack (overwrites the stacks) | Software attack on the buffer with User-to-network channel ( UDP, port 1434 ) | Controllable request |

4.  ***Morris Attack classification:*** The Worm was also popular by some other name such as the Great Worm and the Internet Worm. Cost around $1m. It is counted as the first Internet worm developed by Robert Morris on November 2, 1988 [34]. Language: C; Platform: BSD and SunOS; Origin: Cornell, released at MIT. Detail classification of Morris attack with proposed and other attack classification are shown in Table.5.

TABLE 5
Morris attack taxonomy.

| Hansman with Hunt's grouping [10] | | | | | |
|---|---|---|---|---|---|
| Threat Name | First Dimension | Second Dimension | Third Dimension | Fourth Dimension | |
| Morris | Computer network-based virus | BSD four (4) and Sun three (3) and VAX options | Design and setup for implementation | Facility stealing and subdivision | |
| Joshi's ADMIT Classification [31] | | | | | |
| Attack Name | Attack Vector | Defence | Method | Impact | Target |
| Morris | Misconfiguration | Internet file checking | Internet Worm | Distort | BSD, SunOS |
| Proposed Taxonomy: Sequential Question | | | | | |
| Attack Name | Who | Where | How | What | |
| Morris | Joker (Robert Morris, Jr.,Cornell University) | Multiple PC and Host | Software Attack with the network to the network channel | Controllable request | |

5.  ***MS Remote Procedure Call Attack classification:*** A service by windows server named as Remote Procedure Call (RPC) mostly caused a buffer overflow in system spreads in 2008. Also known as "in the wild" [35]. Type: Internet Worm. Date of published: October 23, 2008. Source language: Assembly. Platform: MS Windows 2000, XP and Server 2003. Detail classification of MS RPC with proposed and other attack classification are shown in Table.6.

TABLE 6
MS RPC attack taxonomy.

| Classification by Lough: VERDICT [8] | | | | | |
|---|---|---|---|---|---|
| Threat Name | Inappropriate Validation | Inappropriate Exposure | Improper Randomness | Inappropriate Reallocation | |
| MS RPC Stack Overflow | None (X) | None (X) | | | |
| Taxonomy by Howard [9] | | | | | |
| Threat Name | Attack utensils, tool | Threat Weakness | Action (way of threat) | Attack Goal | Unlawful outcome |
| MS RPC Stack Overflow | Attack Script | Design | Modify | Process | Increased Access |
| Classification by Hansman with Hunt [10] | | | | | |
| Threat Name | First Dimension | Second Dimension | Third Dimension | Fourth Dimension | |
| Microsoft RPC overflow of the stack | Stack run-off buffer | Microsoft Windows Server Computer | CVE-2008-4250 | Data tampering | |
| AVOIDIT Taxonomy [29] | | | | | |
| Threat Name | Attack Vector (AV) | Operational Impact (OI) | Defence (D) | Impact (I) | Target (T) |
| MS RPC Stack run-off | System buffer run an out-off stack | Installed Malware: ACE | Distort | Solution: RA VU#827267 Solution: a patch of the system | Operating system: MS Server |
| Proposed Taxonomy: Sequential Question | | | | | |
| Attack Name | Who | Where | How | What | |
| Microsoft RPC Overflow of the stack buffer | Little sisters | Group of PC attacked from segment-to-segment based | Software attack via User to network channel (Oversized request) | Traffic volume and Controllable request | |

Overall, From Table. 2 to Table. 6 assessment of 'Sequential Question' classification is done by relating with few other noticeable classifications. The conclusion of the comparisons concludes three drawbacks of previous classifications. Firstly, useful information was unavailable when Lough describes the threat with VERDICT. Secondly, classification done by Howard's helps us only with common evidence. Thirdly, valuable information about the method of payload was supplied by Hunt and Hangman's classification. It also provides little information regarding target, operation and vulnerability but no particular preventive measures were mentioned.

A study by Aziz proposed prevention according to attack classification [36]. Another study also mentioned other protections mechanism against security issues related to fingerprint forgery [37]. A collective anomaly detection techniques were analysed in this study, where data from network traffic were taken into care [38-39]. Similarly, the proposed 'Sequential Question' classification can deliver data to a system administrator regarding the attackers, also the technique of attack, threat consequence to decrease attack's influence. Probable defence mechanism by proposed mechanism can be as shown in Table. 7:

TABLE 7
Defence actions based on Sequential question (proposed) taxonomy.

| Sequential Question | Defence Action |
|---|---|
| WHO | The attacker and his intention are known to the administrator. Therefore management and administrator can work as below:<br>1. He/she can take action against attacker after that secure system<br>2. Secure system and thanks for identifying vulnerability<br>3. International meeting and resolve<br>4. Just secure system and save system |
| WHERE | Attacked source and way is known. Therefore administrator can install filtering systems like firewalls, spam filters, censorware and wiretaps. Certain system and devices can be marked as risky for easy identification and recovery. |
| HOW | Administrator knows through which it will be affected. If the administrator knows how the system will be affected he can take extra care or isolate those parts for extra care. |
| WHAT | Finally, if the administrator knows the characteristics of the computer system after attack then it is easy to decide the safety level that should be installed. The virus can partially attack a certain system or take control of the whole system after a successful attack. The administrator can act according to the type of network attack. Avoidance of result from the particular system can be a possible preventive measure. |

## 5. Conclusion:

This taxonomy is completely different approach than existing attack classification practices which is based on consecutive questions. Our methodology identified common queries towards a threat to detect detail behaviour then classify accordingly. With these four questions (Who, Where, How and what), a network attack type can be clearly acknowledged. A detail validation of this work and comparison with other works has delivered for validation and justification. Our study has successfully identified in detail classification for risky threats like blaster, Melissa, Slammer, MS RPC Stack Overflow and Morris. Proposed classification also satisfied most of the requirements needed for development of an attack taxonomy.

From where the attack is being initiated, to whom it will attack, to what extent the attack affected the system can easily be identified if the answers to those sequential questions are correctly found. However, if the type and gravity of aforementioned attacks could be easily identified far ahead, we can yield a protection mechanism based on that information. It has been observed that threat can be initiated by either a professional hacker or humorous attacker. It has also been perceived that source of the attack can be from solo or in a group. In the same way, the proposed method detected that threat can spread from multiple sources like a computer, flash drive, wire, WiFi etc. Lastly, proposed taxonomy detected that consequence of attack can either be a takeover of the system control or just affect without sharing to other connected systems.

Eventually, our future goal is to construct a taxonomy based on the correlation among personal identification information with the publicly available Internet of Things (IoT) data. Since privacy of personal data is crucial for protecting individuals' credentials, our aim is to define a classification based on their vulnerability level to create a priority among easily available IoT data. In the era of industry 4.0, everything will be connected and rate of privacy breaching will also increase. However, currently, we generalized every kind of data protection with the same level of security measures which is quite insecure for future days. Therefore, the future objective is to generate a taxonomy based on a number of correlated IoT information needed to identify an individuals' identity.

Overall, the safety of the system depends fully on attack detection and subsequent security mechanism. This indicates the necessity of a taxonomy which provides detail information about an attack. Obviously, proposed progressive question centred attack taxonomy can squeeze attacks to extract detail information for assisting system administrators.

**Acknowledgements:** Our gratitude to the C&C research laboratory associates for technical backing. We appreciate Nuzhat Mariam, Dong-A University for doing overall grammatical improvement.

## References

[1]  Cohen, F. (1997). Information system attacks: A preliminary classification scheme. *Computers & Security*, *16*(1), 29-46.

[2]  Landwehr, C. E., Bull, A. R., McDermott, J. P., & Choi, W. S. (1993). *A taxonomy of computer program security flaws, with examples* (No. NRL/FR/5542--93-9591). NAVAL RESEARCH LAB WASHINGTON DC.

[3]  Bishop, M. (1995). *A taxonomy of unix system and network vulnerabilities*. Technical Report CSE-95-10, Department of Computer Science, University of California at Davis.

[4]  Rae, A., & Wildman, L. (2003). A taxonomy of attacks on secure devices. In *Proceedings of the Australia Information Warfare and Security Conference 2003* (pp. 251-264). York.

[5]  Álvarez, G., & Petrović, S. (2003). A new taxonomy of web attacks suitable for efficient encoding. *Computers & Security*, *22*(5), 435-449.

[6]  Wood, A. D., & Stankovic, J. A. (2004). A taxonomy for denial-of-service attacks in wireless sensor networks. *Handbook of Sensor Networks: Compact Wireless and Wired Sensing Systems*, 739-763.

[7]  Mirkovic, J., & Reiher, P. (2004). A taxonomy of DDoS attack and DDoS defense mechanisms. *ACM SIGCOMM Computer Communication Review*, *34*(2), 39-53.

[8]  Lough, D. L. (2001). *A taxonomy of computer attacks with applications to wireless networks* (Doctoral dissertation, Virginia Tech).

[9]  Howard, J. D., & Longstaff, T. A. (1998). A common language for computer security incidents (No. SAND98-8667). Sandia National Labs., Albuquerque, NM (US); Sandia National Labs., Livermore, CA (US).

[10]  Hansman, S., & Hunt, R. (2005). A taxonomy of network and computer attacks. *Computers & Security*, *24*(1), 31-43.

[11]  Lippmann, R., Haines, J. W., Fried, D. J., Korba, J., & Das, K. (2000). The 1999 DARPA off-line intrusion detection evaluation. *Computer networks*, *34*(4), 579-595.

[12]  Lindqvist, U., & Jonsson, E. (1997, May). How to systematically classify computer security intrusions. In *Security and Privacy, 1997. Proceedings., 1997 IEEE Symposium on* (pp. 154-163). IEEE.

[13]  Bisbey, R., & Hollingworth, D. (1978). Protection analysis: Final report. *ISI/SR-78-13, Information Sciences Inst*, 3.

[14]  Abbott, R. P., Chin, J. S. (1976). *Security analysis and enhancements of computer operating systems*. NATIONAL BUREAU OF STANDARDS WASHINGTONDC INST FOR COMPUTER SCIENCES AND TECHNOLOGY.

[15]  Abrek, N. (2015). Attack Taxonomies and Ontologies. *Future Internet 2014, Network Architectures and Services*.

[16]  Weaver, N., Paxson, V., Staniford, S., & Cunningham, R. (2003, October). A taxonomy of computer worms. In *Proceedings of the 2003 ACM workshop on Rapid malcode* (pp. 11-18). ACM.

[17]  Ning, C., Chen, M., & Zhou, D. (2014). Hidden Markov model-based statistics pattern analysis for multimode process monitoring: an index-switching scheme. *Industrial & Engineering Chemistry Research*, *53*(27), 11084-11095.

[18] Ahmed, M., & Ullah, A. S. B. (2017). False Data Injection Attacks in Healthcare. 15th Australasian Data Mining Conference, Melbourne, Australia. Conferences in Research and Practice in Information Technology, Vol. 172.

[19] Onik, M. M. H., Al-Zaben, N., Phan Hoo, H., & Kim, C. S. (2017). MUXER—A New Equipment for Energy Saving in Ethernet. Technologies, 5(4), 74.

[20] Bhuyan, M. H., Bhattacharyya, D. K., & Kalita, J. K. (2017). Network Traffic Anomaly Detection Techniques and Systems. In Network Traffic Anomaly Detection and Prevention (pp. 115-169). Springer, Cham.

[21] Lippmann, R. P., & Cunningham, R. K. (2000). Improving intrusion detection performance using keyword selection and neural networks. *Computer networks*, *34*(4), 597-603.

[22] Don Stikvoort. Incident Classifcation / Incident Taxonomy according to eCSIRT.net adapted.2012.url:https://www.trustedintroducer.org/Incident Classification Taxonomy.pdf (visited on 17/08/2017).

[23] Daigle, R. C. (2001). *An Analysis of the Computer and Network Attack Taxonomy* (No. AFIT/GIR/ENV/01M-04). AIR FORCE INST OF TECH WRIGHT-PATTERSONAFB OH.

[24] Hollingworth, D. (2003). Towards threat, attack, and vulnerability taxonomies. In *IFIP WG* (Vol. 10).

[25] Kumar, M. (2012, September 10). The 10 Most Infamous Student Hackers of All Time. Retrieved February 14, 2018, from https://thehackernews.com/2012/09/the-10-most-infamous-student-hackers-of.html

[26] White hat (computer security). (2018, February 14). Retrieved February 14, 2018, from https://en.wikipedia.org/wiki/White_hat_(computer_security)

[27] Conger, K. (2017, July 25). The Politics of Hacking in the Age of Trump. Retrieved February 14, 2018, from https://gizmodo.com/the-politics-of-hacking-in-the-age-of-trump-1797188587

[28] Van Heerden, J. (1979). The morphology and taxonomy of Euskelosaurus from South Africa. Nasionale Museum.

[29] Simmons, C., Ellis, C., Shiva, S., Dasgupta, D., & Wu, Q. (2009). AVOIDIT: A cyber attack taxonomy. In Proc. of 9th Annual Symposium On Information Assurance-ASIA (Vol. 14).

[30] Leyden, J. (2003, August 14). Blaster rewrites Windows worm rules. Retrieved February 14, 2018, from http://www.theregister.co.uk/2003/08/14/blaster_rewrites_windows_worm_rules/

[31] Joshi, C., & Singh, U. K. (2014). Admit-A five dimensional approach towards standardization of network and computer attack taxonomies. International Journal of Computer Applications, 100(5).

[32] Melissa Macro Virus. (1999, March 27). Retrieved February 14, 2018, from https://www.cert.org/historical/advisories/CA-1999-04.cfm

[33] Moore, D., Paxson, V., Savage, S.(2003). Inside the slammer worm. IEEE Security & Privacy, 99(4), 33-39.

[34] *Brendan P. Kehoe (2007). "The Robert Morris Internet Worm". mit. Retrieved* February 14, 2018*, from http://groups.csail.mit.edu/mac/classes/6.805/articles/morris-worm.html*

[35] Pincus, J., & Baker, B. (2004). Beyond stack smashing: Recent advances in exploiting buffer overruns. *IEEE Security & Privacy*, *2*(4), 20-27.

[36] Aziz, A., Lai, W. L., & Manni.(2014). U.S. Patent No. 8,898,788. Washington, DC: U.S. Patent and Trademark Office.

[37] Chaudhry, J., Qidwai, U., Miraz, M. H., Ibrahim, A. (2017). Data security among ISO/IEEE 11073 compliant personal healthcare devices through statistical fingerprinting. Institute of Electrical and Electronics Engineers.

[38] Ahmed, M. (2018). Collective Anomaly Detection Techniques for Network Traffic Analysis. *Annals of Data Science*, 1-16.

[39] Islam, M. M., Hassan, M. M., Alamri, A., & Huh, E. N. (2016). Data classification and scheduling for sensor virtualisation scheme in public healthcare system. *International Journal of Sensor Networks*, *22*(4), 259-273.